\documentclass[twocolumn,english,aps,prb,superscriptaddress,natbib,bibnotes,amsmath,amssymb,floatfix,groupedaddress,footinbib]{revtex4-2}

\usepackage[colorlinks=true, allcolors=blue]{hyperref}

\usepackage[utf8]{inputenc}
\usepackage[english]{babel}

\usepackage{amsmath,amsfonts,amssymb}
\usepackage{bbold}
\usepackage{mathtools}
\usepackage[T1]{fontenc}
\usepackage{url}

\usepackage{amsmath}
\usepackage{siunitx}

\usepackage{amsfonts}
\usepackage{amssymb}

\usepackage{graphicx}
\graphicspath{{./figures/}}

\begin{document}

\title{Post-selection free time-bin entanglement on a thin-film lithium niobate photonic chip}

\author{Marcello Bacchi$^{1,\dag,*}$, Andrea Bernardi$^{2,3,\dag}$, Marco Clementi$^1$, Sara Congia$^{1,4}$, Francesco Garrisi$^3$, Andrea Martellosio$^3$, Marco Passoni$^3$, Alexander Wrobel$^3$, Federico Andrea Sabattoli$^3$, Matteo Galli$^1$, Daniele Bajoni$^2$
}

\affiliation{
$^1$Dipartimento di Fisica “A. Volta”, Università di Pavia, Via Bassi 6, 27100 Pavia, Italy.\\
$^2$Dipartimento di Ingegneria Industriale e dell’Informazione, Università di Pavia, Via Ferrata 5, 27100 Pavia, Italy.\\
$^3$Advanced Fiber Resources Milan s.r.l., Via Fellini 4, 20097 San Donato Milanese, Italy.\\
$^4$Université Grenoble Alpes, CEA-Leti, 38054 Grenoble, France.
}

\date{\today}

\begin{abstract}
\noindent
Time-bin entanglement is the most commonly used form of entanglement for quantum communication protocols over fiber networks, due to the natural resilience of this encoding scheme to thermal phase fluctuations in optical fibers. 
Projective measurements on some bases in the time-bin encoding need, however, post-selection of the measured events, introducing a loophole in Bell tests and requiring high temporal resolution.
In this work, we demonstrate chip-integrated receivers for time-bin entanglement certification including a high-speed optical switch to remove such post-selection loophole. 
The receivers are realized using thin-film lithium niobate and operate at a switching frequency of 5 GHz, enabling high secure key rates.
We demonstrate a Bell inequality violation by more than 24 standard deviations without the need for time resolution at the time-bin separation level.
\end{abstract}

\maketitle

\section{Introduction}
\noindent
Entangled states of light play a fundamental role in quantum information science and technology\cite{Moody2022}, with applications ranging from fundamental tests of quantum mechanics to quantum computing and quantum communication.
Their use in conjunction with fiber optic networks makes them a cornerstone of quantum communication protocols, where they are exploited in protocols such as quantum teleportation\cite{Bennett1993}, entanglement swapping\cite{Llewellyn2020}, and entanglement-based quantum key distribution (QKD)\cite{Xu2020}. 
In this context, the encoding of quantum information in the discrete time basis, i.e. the time-bin (TB) encoding\cite{Brendel1999, Marcikic2002}, has proven to be particularly favorable in field tests compared to other encoding schemes, such as path or polarization, especially over fiber\cite{Inagaki:13, kim:22} and free space-links\cite{vallone:16, Jin:19, cocchi:25}.
Key advantages of TB-entangled states include their robustness against noise over large distances, ease of generation and manipulation, and the possibility to scale the data rates through hyper-entanglement with other degrees of freedom \cite{chapman22, Zhong:24, stein:16} or by increasing the dimension of the Hilbert space (TB \textit{qudits})\cite{Yu2025}.

% The post-selection loophole
The correlated and non-local nature of the TB state is ensured by Bell's theorem and can be assessed through a modified version of the well-known Franson interferometer \cite{Franson1989}, originally proposed by Brendel \textit{et al.}\cite{Brendel1999}.
On the other hand, two-photon interference in the TB basis only occurs only on a subset of the detected photon pairs, associated with indistinguishability of the interferometric path experienced.
As early pointed out by Aerts \textit{et al.}\cite{Aerts1999}, this opens a post-selection loophole (PSL) that affects local-realistic tests of quantum mechanics, hindering the security of QKD protocols.
The need for post-selection also requires the use of detectors with high temporal resolution, capable of distinguishing between the time bins.
This poses a trade-off in which larger time separations are desirable to distinguish the time bins, but these in turn entail limited bandwidth of the quantum communication protocols and long delay lines in the analyzing interferometers to perform projective measurements. 
This also results in the need for active feedback loops to stabilize the interferometers. .
Strategies proposed to overcome temporal post-selection employed hyperentangled states combining the polarization and TB degrees of freedom, as demonstrated by Strekalov \textit{et al.}\cite{Strekalov1996}, albeit at the expense of additional complexity.
Another solution is the so-called “hug” interferometric configuration, as proposed by Cabello \textit{et al.}\cite{Cabello2009}, although this approach requires active stabilization of the optical link.
A third solution, first proposed by Vedovato and coworkers\cite{Vedovato2018}, leverages active switching to remove the need for post-selection.
Unlike in common Franson-type interferometry, here the TB paths are \textit{deterministically} selected by active switches to grant two-photon interference for virtually any pair of photons detected, closing the PSL vulnerability and overcoming the need to resolve and discard detection events.
In its first demonstration, this approach used a pair of fiber-based actively stabilized Mach-Zehnder interferometers, whose length matched the separation of the TB pairs, combined with commercial electro-optic modulators used as fast active switches.

All of the approaches enumerated above have shown through proof-of-concept demonstrations how TB entanglement can still be regarded as a useful resource for quantum communication, although at the expense of the additional complexity required to overcome temporal post-selection.
However, to make TB entanglement scalable and practical, integrated photonics provides a promising avenue to reduce complexity, size, and cost while maintaining performance.
A demonstration of the hug scheme, for example, has been recently reported on a silicon photonic chip\cite{Santagiustina2024}.

% Overview of non TFLN
Recently, the thin-film lithium niobate (TFLN) integrated photonic platform has emerged as a major player in the photonics technology landscape\cite{Zhu2021}.
Combining the high second-order nonlinearity of this material with tight confinement and low propagation losses in etched waveguides, TFLN photonic integrated circuits (PICs) enable the implementation of many nonlinear devices on the same photonic chip with high nonlinear efficiency and low bending losses.
With regard to integrated quantum photonics, TFLN has proven suitable for nonclassical light generation on-chip through spontaneous parametric down-conversion (SPDC) \cite{Zhao2020, Nehra2022, Xin2022} and its manipulation \cite{Stokowski2023, Park2024, Finco2024}, unattainable in centrosymmetric media such as silicon and silicon nitride, and for the high-speed modulation thanks to its high Pockels coefficient\cite{Zhu2022, Sund2023}.

% Overview of the work
In this work, we leverage the industrial-grade TFLN photonic platform developed by Advanced Fiber Resources Milan, to develop a turnkey device for the certification of post-selection-free TB entanglement.
The device takes advantage of the $\chi^{(2)}$ properties of TFLN to achieve high-speed ($\SI{5}{\giga\hertz}$) active switching on-chip, effectively implementing the approach outlined in Ref.\cite{Vedovato2018} in a centimeter-scale device, fully packaged and interconnected optically and electrically, overcoming the need for actively stabilized interferometers and bulky optical components.
We benchmark the device by showing genuine TB entanglement of photon pairs generated by a standard on-chip spontaneous four-wave mixing source pumped with gigahertz-level repetition rates, with a pulse separation lower or comparable with the detector jitter.
Our results confirm a violation of the Bell inequality by more than 24 standard deviations without temporal post-selection and subtraction of accidentals, showing how this approach allows for the certification of genuine TB entanglement even with low-resolution detectors.
\begin{figure}[t]
    \centering
    \includegraphics[width=7cm]{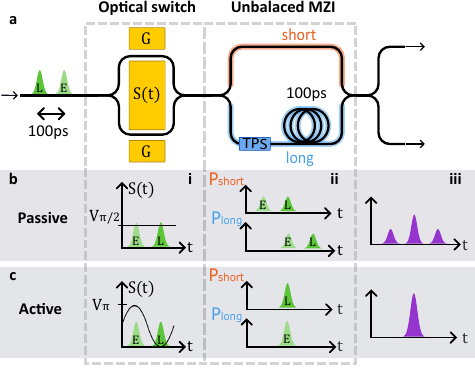}
    \caption{(a) Schematic of the TFLN integrated photonic circuit. TPS: Thermal Phase Shifter, S: RF Signal, G: Ground. A pair of laser pulses, separated by \SI{100}{\pico\second} are injected in the device. 
    (b-c) Evolution of the input laser pulses in the integrated device for both passive (b) and active (c) configurations: 
    i) phase applied to the input state in the MZM. 
    ii)  Evolution after propagation in the unbalanced MZI, before recombination. 
    iii) Light detection pattern at the device output.}
    \label{fig:TFLN_schematic}
\end{figure}

\section{\label{sec:theory}Theory and operating principle}
\noindent
We first describe the evolution of generic light pulses within the TFLN receiver, in order to straightforwardly outline the device operating principle.
Figure~\ref{fig:TFLN_schematic}a depicts a scheme of the TFLN device: pairs of early ($E$) and late ($L$) pulses are fed at the input of the device, pass through a balanced Mach-Zehnder modulator (MZM), or “optical switch”, and then through an unbalanced Mach-Zehnder interferometer (MZI). 
When operating the device in a “passive” configuration, namely in the absence of radio-frequency (RF) modulation, the optical switch is set to operate as a 50:50 beam-splitter (Fig.~\ref{fig:TFLN_schematic}b i), randomly routing the incoming photons to the long or short arm of the unbalanced interferometer. Pulses traveling the long arm are delayed by $T_\mathrm{s}=\SI{100}{\pico\second}$, where $T_\mathrm{s}$ is the optical path separation (Fig.~\ref{fig:TFLN_schematic}b ii), resulting in the characteristic three-pulse pattern at detectors after recombination (Fig.~\ref{fig:TFLN_schematic}b iii). Interference arises only at the central measured peak, where two pulses temporally overlap. Conversely, when the device operates in an “active” configuration, the optical switch is biased at the quadrature point and fed with an electrical RF signal at frequency $f_\mathrm{m}=1/2T_\mathrm{s}$ and peak-to-peak modulation amplitude $V_\pi$ (Fig.~\ref{fig:TFLN_schematic}c i). Here, the early and late pulses are deterministically routed to the long and short arms of the MZI (Fig.~\ref{fig:TFLN_schematic}c ii), respectively. This configuration allows for the complete temporal overlap of the pulses after recombination (Fig.~\ref{fig:TFLN_schematic}c iii), thus the totality of input light is subject to interference.

\begin{figure*}
    \centering
    \includegraphics[width=\linewidth]{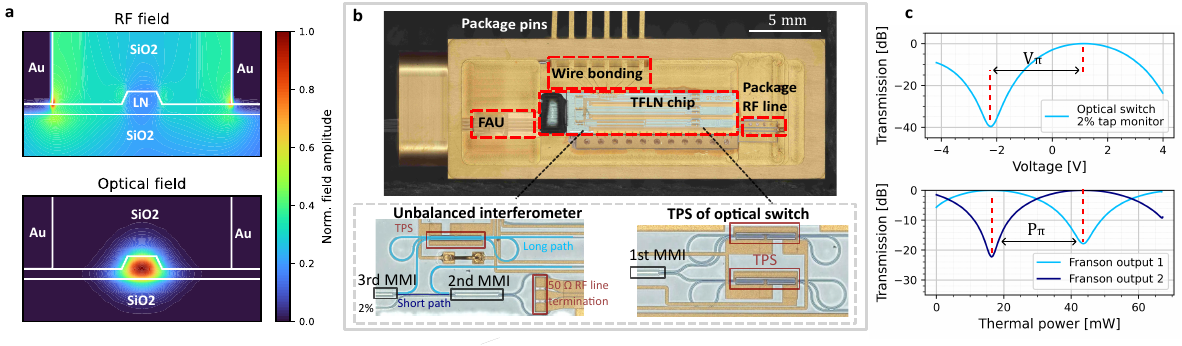}
    \caption{
    (a) Simulated transverse component (parallel to the extraordinary optical axis) of the RF and optical electric field in the MZM cross-section.
    (b) Micrograph of the packaged device. FAU: Fiber Array Unit.
    The insets show details of the main components of the PIC.
    (c) Linear characterization. Upper panel: balanced MZM used as optical switch, $V_\pi$ of \SI{3.37}{\volt}. Lower panel: unbalanced MZI, $P_\pi$ of \SI{27.3}{\milli\watt}.
    }
    \label{fig:TFLN_image}
\end{figure*}

So far, we have considered generic (classical) light pulses, i.e. temporally separated wavepackets. We now focus on the evolution of correlated photon pairs: in particular, we are interested in TB-entangled bi-photon states, which can be generated injecting light pulses in a nonlinear medium through parametric fluorescence processes such as spontaneous four-wave mixing (SFWM) or SPDC and then used as input to the device. Selecting a time separation between twin pulses much higher than the pulse temporal width and minimizing multi-pair emission from the crystal, the resulting state can be described as\cite{Brendel1999}:
\begin{equation}
    |\Phi\rangle = \frac{1}{\sqrt{2}} \left(|E\rangle_\mathrm{s} |E\rangle_\mathrm{i} + e^{i \varphi_\mathrm{p}} |L\rangle_\mathrm{s} |L\rangle_\mathrm{i} \right) ,
\end{equation}
where $\varphi_\mathrm{p}$ is a relative phase inherited from the pump pulses, while $|E\rangle_\mathrm{s (i)}$ and $ |L\rangle_\mathrm{s (i)}$ describe the emission of a signal (idler) photon either in the early or late pump pulse wavepacket, respectively.

In the “passive” configuration, signal and idler photons can be detected at three different time delays with respect to the pump trigger signal, yielding the characteristic three-peak delay histogram profile. Post-selecting the central coincidence peak, the projection $\hat{P}_{\mathrm{s,i}} = |\psi\rangle\langle\psi|_{\mathrm{s,i}}$ is realized, where $|\psi\rangle_{\mathrm{s,i}} = \left( |E\rangle_{\mathrm{s,i}} \pm e^{i \varphi_{\mathrm{s,i}}}|L\rangle_{\mathrm{s,i}} \right)/\sqrt{2}$. 
$\varphi_{\mathrm{s,i}}$ are phase shifts that originate from analysis interferometers, i.e. the unbalanced MZI, and controlled by thermal phase shifter (Fig.~\ref{fig:TFLN_schematic}a).  
The probability of coincidence detection around the central peak is
\begin{equation}
\label{eq:post_sel_probability}
    \mathcal{P}\left(\varphi_\mathrm{s},\varphi_\mathrm{i}\right) = \frac{1}{4}\left[1\pm\mathcal{V}_\mathrm{ps} \cos\left(\varphi_\mathrm{s}+\varphi_\mathrm{i}-\varphi_ \mathrm{p}\right)\right]  ,
\end{equation}
where $\mathcal{V}_\mathrm{ps}$ is the post-selected visibility of the two-photon interference fringe. The maximum value of the parameter $S$ of the Clauser-Horn-Shimony-Holt (CHSH) inequality is $S_\mathrm{max} = 2\sqrt{2}\mathcal{V}$, hence the CHSH inequality $\left(S\leq 2\right)$ is violated if $\mathcal{V}>1/\sqrt{2}$. Without central peak post-selection, thus considering the whole coincidence pattern, coincidence detection probability is still similar to Eq. \ref{eq:post_sel_probability}, while the global visibility becomes $\mathcal{V}_g = \mathcal{V}_\mathrm{ps}/4$. It results that CHSH inequality can not be violated, as the maximum achievable visibility is 25\% \cite{Vedovato2018}.

Activating the optical switch, i.e. feeding the MZM with a synchronous electrical RF signal, the early and late TBs are deterministically routed to the long and short arms of the second interferometer, respectively. 
In this case, the observer implements the positive operator valued measure (POVM):
\begin{equation}
    \hat{\Pi}_{\mathrm{s,i}} = \frac{1}{2} \cos^2 \left(\frac{\varphi_E}{2}\right) \mathbb{1} + \sin^2 \left(\frac{\varphi_E}{2}\right) \hat{P}_{\mathrm{s,i}} , 
\end{equation}
where $\mathbb{1} = |E\rangle\langle E| + |L\rangle\langle L|$ and $\hat{P}_{\mathrm{s,i}}$ was defined earlier. Imposing $\varphi_E = \pi$ results in $\hat{\Pi}_{\mathrm{s,i}} = \hat{P}_{\mathrm{s,i}}$ and a single peak delay histogram is retrieved without data post-selection. Data shedding is no longer necessary to exhibit quantum interference, since the CHSH inequality is violated freed from the post-selection loophole.

\begin{figure*}
    \centering
    \includegraphics[width=\linewidth]{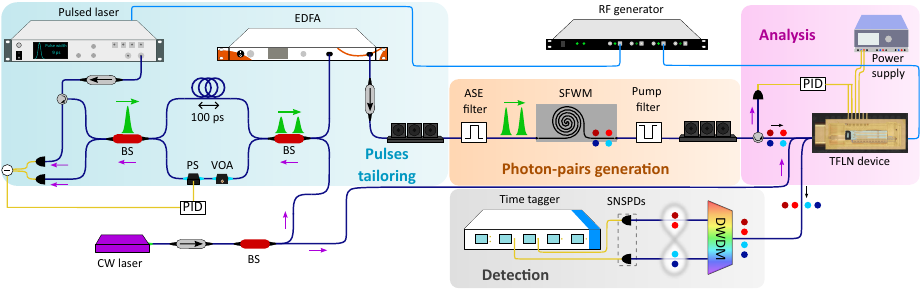}
    \caption{Experimental setup: a mode-locked laser (Pritel) emits pulses at \SI{1550.92}{\nano\meter}. 
    Each pulse is delayed by an unbalanced fiber MZI to create the early/late pair and amplified by an EDFA (Keopsys). 
    An in-fiber phase shifter, driven by a PID, controls the phase difference between the two arms of the MZI and a variable optical attenuator (VOA) equalizes the power of the pulses. 
    After filtering of the amplified spontaneous emission (ASE) the tailored pump pulse pairs are coupled to a silicon waveguide, where photon pairs are generated via SFWM. 
    The residual pump is then filtered and the generated photons are routed to the TLFN device for entanglement certification. 
    Here, the unbalanced interferometer phase is controlled by a TPS, driven by an external power supply, while the optical switch is actuated by the same RF generator driving the pump laser.
    Light is demultiplexed with in-fiber filters and finally detected by SNSPDs (Photon Spot). 
    A CW laser (PurePhotonics) counter-propagates in the first delay line and in the device MZM to stabilize the operating point.}
    \label{fig:setup}
\end{figure*}

\section{\label{sec:device}Device design and linear characterization}
\noindent
The integrated device was fabricated at Advanced Fiber Resources Milan on a 4-inch lithium-niobate-on-insulator (LNOI) X-cut wafer with silicon substrate.
The optical waveguides are defined in a \SI{550}{\nano\meter}-thick LN layer via a \SI{275}{\nano\meter}-deep etching process and are embedded in a silica matrix. They ensure fundamental mode propagation in the standard C+L optical telecom bands, with optical propagation losses around \SI[per-mode = symbol]{0.2}{\decibel\per\centi\meter}.
Fig.~\ref{fig:TFLN_image}b illustrates a microscope image of a fabricated device: the input photons are coupled through one of the four edge couplers and pass through a 50:50 multimode interferometer (MMI) to form the balanced MZM, that is used as optical switch. 
In the MZM, \SI{1}{\centi\meter}-long electrodes form a ground-signal-ground coplanar RF line that applies the electro-optic (EO) modulation to the two optical waveguides in a push-pull configuration. 
The two optical paths of the MZM are geometrically balanced and each one has a thermal phase shifter (TPS) to control the MZM bias. 
The optical switch is completed by a 2×2 MMI, whose two outputs form an unbalanced interferometer with a fixed delay of \SI{100}{\pico\second}. 
The unbalanced interferometer phase can be controlled in the longer path via either a TPS or an EO phase modulator independently. 
Finally, it is closed by a third 2×2 MMI, where the two complementary outputs are routed to two edge couplers, which provide convey light out of the chip.

The balanced MZM is the key component of our integrated device, as it is designed for high-speed single photon switching. 
The optical switch routes the early (late) photon to the longer (shorter) path of the unbalanced interferometer, thus eliminating the need of temporal post-selection.
To do this, the optical switch must apply a phase difference of $\pi$ within the temporal separation of the TB state (Fig.~\ref{fig:TFLN_schematic}c), that is \SI{100}{\pico\second}, which means applying a modulation with a fundamental frequency of the switch at \SI{5}{\giga\hertz}. 
Therefore, the electrodes geometry that defines the RF line must be carefully designed to optimize the optical switch performance. 
Key design considerations include: (i) tailoring the RF line characteristic impedance to match the the external signal generator impedance of \SI{50}{\ohm} for maximum power transfer, (ii) matching the RF phase velocity with the optical group velocity to maximize the EO modulation bandwidth, and (iii) optimizing the RF and optical field overlap to minimize $V_{\pi}$ (Fig.~\ref{fig:TFLN_image}a).

Figure~\ref{fig:TFLN_image}c displays some illustrative data from the linear characterization of the device: $V_{\pi} = \SI{3.37}{\volt}$ at low frequency (\SI{1}{\mega\hertz}) and a simulated \SI{3}{\decibel} EO bandwidth of \SI{21.4}{\giga\hertz} (upper panel). The tested TPS $\pi$ phase difference is $P_{\pi} = \SI{27.3}{\milli\watt}$ for the TPS (lower panel).

To guarantee the portability of the proposed device, the fabricated TFLN chip is fully packaged in a metal case, as shown in Fig.~\ref{fig:TFLN_image}b. 
The packaging includes a 4-fiber (Nufern UHNA7) array unit (FAU) pigtailed to the chip at maximum optical coupling. 
The chip is attached to the package using a thermally conductive adhesive to improve heat dissipation and thermal stabilization. 
A standard G3PO RF connector and precise wire bonding between the package and the chip feed the modulation signal to the MZM electrodes. 
All the TPSs and the EO electrodes of the unbalanced interferometer are connected to lateral package pins by wire bonding with the relative chip pads. 

\begin{figure*}[t]
    \centering
    \includegraphics[width=\linewidth]{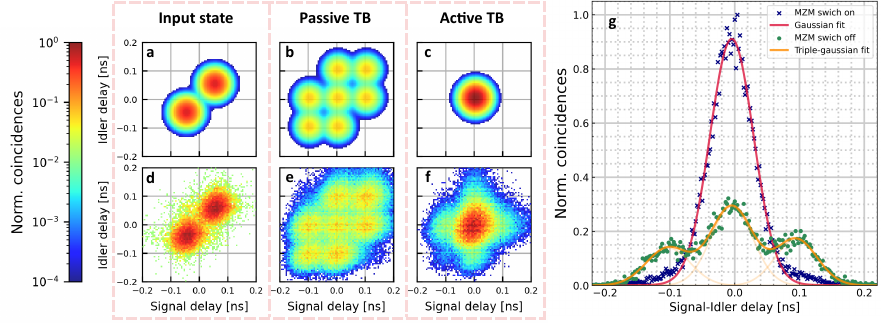} 
        \caption{(a-c) Simulated and (d-f) reconstructed JTIs for the input state and after undergoing evolution in the device, both in the passive and active switching configurations.
        The amplitude of the passive switching case is normalized to the peak of the active switching one for consistency.
        The peaks resolution is limited by SNSPDs jitter (\SI{50}{\pico\second}) and the pulse delay is \SI{100}{\pico\second}.  
        (g) Experimental TCSPC coincidence histograms for the passive and active switching cases. 
        The same data can in principle be retrieved by integrating along the main diagonal of the JTIs.
        }
            \label{fig:JTI_hist}
\end{figure*}

\section{\label{sec:experiment}Experiment}

\noindent
To demonstrate the device operating principle and assess its performance, we generate a TB-entangled state using an integrated SFWM source and route both signal and idler photons to one of the fabricated devices, where we perform measurements both in passive and active configurations.
The experimental setup is depicted in Fig.~\ref{fig:setup}. 
An actively mode-locked, infrared laser emits pump pulses \SI{9}{\pico\second} wide at \SI{1550}{\nano\meter} wavelength, with a repetition rate of \SI{500}{\mega\hertz}. 
The pulses pass through an optical isolator and are sent to an unbalanced MZI, with a free-space time delay that matches the \SI{100}{\pico\second} of the on-chip interferometer (light blue panel). 
Pulses are then amplified with an erbium-doped fiber amplifier (EDFA) and coupled to a silicon photonic chip, where signal and idler photons are generated via SFWM in a low-loss \SI{16}{\milli\metre}-long silicon-on-insulator rib waveguide (orange panel). 
Photons are collected and, after pump rejection, routed to the TFLN device input to perform Franson interferometry in the passive or active TB configuration. 
The phase difference between short and long interferometer arms is set via an external power supply which drives the TPS (violet panel). 
An RF source generates the modulation signals driving both the pulsed laser and the optical switch, hence they are synchronized. 
Signal and idler photons coming from a single output of the MZI are then demultiplexed and sent to superconducting nanowire single-photon detectors (SNSPDs), with an 85\% efficiency and jitter of \SI{50}{\pico\second} for time-correlated single photon counting (TCSPC) (grey panel).
Both the in-fiber MZI and the integrated balanced MZM are stabilized via a counter-propagating continuous-wave (CW) laser and proportional-integral-derivative (PID) feedback loops. 
The former fixes the phase difference between pump pulses, the latter maintains the MZM at its quadrature working point. 
The optical switch is modulated using a sinusoidal signal with a peak-to-peak voltage of $V_{\pi}= \SI{4.35}{\volt}$ at a frequency of \SI{5}{\giga\hertz}.
The optical losses from silicon waveguide to photodetectors are \SI{18.6}{\decibel} per signal and idler channel, including silicon chip outcoupling $\left(\SI{4}{\decibel}\right)$ and the TFLN device insertion loss (\SI{6.1}{\decibel}). 

To characterize the device operation on the generated quantum state, we first reconstruct the joint temporal intensity (JTI), of the two-photon state at the output of the device.
This quantity, corresponding to the squared amplitude of the complex biphoton wavefunction in the temporal representation, can be experimentally assessed as a two-dimensional coincidence histogram between signal and idler photon arrival times ($t_s$, $t_i$) and the emission time of the pump pulses.
Before entering the device, the JTI of the biphoton state consists of two main lobes, associated with the two entangled time bins and with theoretical Schmidt number $K=2$ (Fig. \ref{fig:JTI_hist}a).
The device then enacts a unitary evolution of the state, that in the passive configuration transforms it by creating, intuitively, four copies of each of such lobes through temporal shift of the original pattern on the horizontal and vertical axes.
The resulting pattern, depicted in Fig. \ref{fig:JTI_hist}b, consists of seven lobes, of which only the central one displaying quantum interference.
A rigorous explanation of the mechanism behind time-bin interference goes beyond the scope of this work and can be found, for example, in Refs.\cite{Marcikic2002, Santagiustina2024}.
The simulated pattern is well reproduced experimentally, as shown in Fig. \ref{fig:JTI_hist}e, confirming that the device correctly performs the unitary transformation associated with such Franson-type interference.
Conversely, when operated in active switching configuration, the device deterministically shifts the two lobes along the main diagonal of the JTI ($t_i=t_s$), resulting in quantum interference for all photon pairs reaching the device.
The experimental trace, shown in Fig. \ref{fig:JTI_hist}f, well replicates the simulation (Fig. \ref{fig:JTI_hist}c), with only few residual counts in the secondary lobes, which we attribute to finite extinction of the MZM stage.

The signal-idler TCSPC histogram (Fig. \ref{fig:JTI_hist}g), representing a measurement of the Glauber cross-correlation function $g_{\rm s,i}^{(2)}(t_s,t_i)$, can be either directly retrieved from the photon arrival times or equivalently obtained from integration of the JTI along the $t_s=t_i$ axis.
This operation effectively projects the measured JTI over the anti-diagonal $t_i=-t_s$, and it is therefore a function of the arrival time delay at the detectors.
As a consequence, the coincidence histogram trace associated with the passive switching (green dots) regime displays three peaks, which can be only partially resolved due to the finite resolution of our SNSPD (\SI{50}{\pico\second}).
Since the central peak originates from the three diagonal lobes, varying the interferometer phase only allows partial extinction. This results in a maximum theoretical visibility of 50\%, when post-selecting only the central peak events, and 25\%, when also including the lateral histogram peaks, preventing PSL-free entanglement certification, as pointed out in Section \ref{sec:theory}.
Conversely, operating the device in active switching regime provides a single-peak coincidence histogram (blue crosses).
Since in this case all the incoming photons participate to quantum interference, the interference can here reach a maximum theoretical visibility of 100\%, therefore allowing a PSL-free violation of Bell's inequality.

We then perform a Bell test by collecting photons from one output channel of the device, without coincidences post-selection, namely summing all the coincidence events associated with the correlation histogram while varying the phase of the unbalanced MZI by acting on the TPS current.

\begin{figure}[t]
    \centering
    \includegraphics[width=8.5cm]{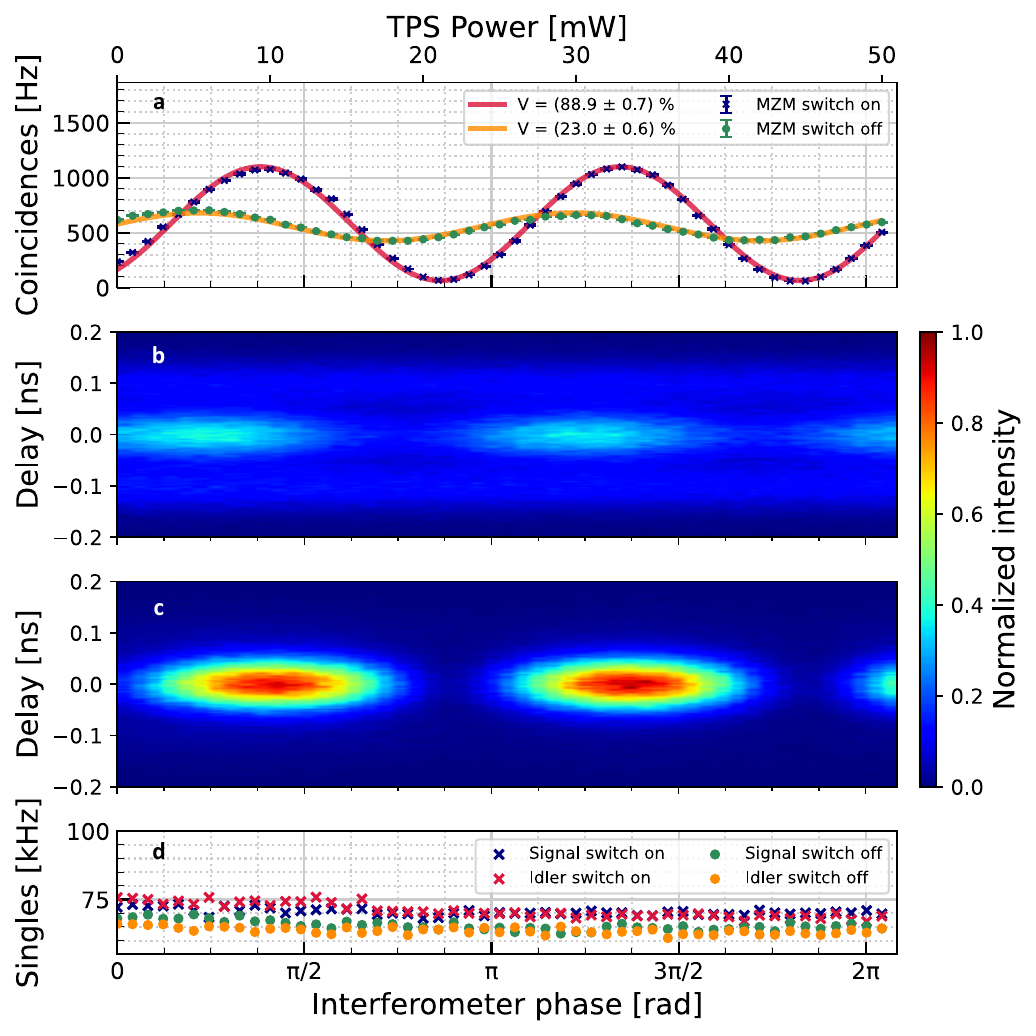}
    \caption{
    (a) Measured quantum interference curves in both passive (MZM off) and active (MZM on) configurations. 
    The solid curves are sinusoidal fits of the experimental data and are used to assess the visibility.
    Experimental intensity plots of the coincidence histograms for:
    (b) the passive scheme, and (c) the active scheme.
    (d) Photon rate at the SNSPDs.
    All the data and visibility are reported without subtraction of accidental counts.}
    \label{fig:bell}
\end{figure}

The result is shown in Fig. \ref{fig:bell}: two different experimental Bell curves, corresponding to the passive (green dots) and active (dark blue crosses) configurations clearly show an increase of the quantum interference visibility between the respective cases.
Sinusoidal fits of the experimental data (each associated with a coincidence histogram, with 20 seconds integration time) yield a visibility of $\left(22.8 \pm 0.7\right)\%$ for the passive case and $\left(87.9 \pm 0.7\right)\%$ for the active case without accidentals subtraction. With accidental correction the visibility are respectively $\left(23.0 \pm 0.6\right)\%$ and $\left(88.9 \pm 0.7\right)\%$.
The latter values exceed the theoretical bound for local realism ($1/\sqrt{2}\approx70.7\%$) by more than 24 standard deviations, proving a PSL-free Bell violation for the TB-entangled state. 

\section{\label{sec:conclusion}Discussion and conclusions}
\noindent
We demonstrated a chip-integrated receiver for the certification of genuine TB entanglement without temporal post-selection.
To the best of our knowledge, this is the first experimental implementation of the method outlined in Ref.\cite{Vedovato2018} using an integrated device.
To this end, we take advantage of an industrial-grade TFLN fabrication technology and develop fully packaged devices that prove the maturity of TFLN technology for the practical deployment of quantum protocols.

Unlike the scheme presented in Ref.\cite{Strekalov1996}, the active switching scheme implemented here does not exploit hyperentanglement or any additional degree of freedom to overcome the PSL. 
Instead, only the TB qubit encoding is required, enabling the exploitation of other degrees of freedom to encode useful quantum information.
By deterministically selecting the long or short path, all photon pairs (excluding losses) can contribute to quantum interference. 
In contrast, the “hug” schemes \cite{Cabello2009, Santagiustina2024} passively discards photons that do not temporally overlap, reducing the number of useful photon pairs by half and limiting the maximum two-photon coincidences.

As shown in Fig. \ref{fig:JTI_hist}f, the sidebands caused by spurious events are almost entirely suppressed during device operation. 
This greatly simplifies the measurement process as the required temporal resolution is now less than the period between two pulses rather than the TB spacing.
We leverage this feature to drive the device at a switching frequency of \SI{5}{\giga\hertz}, among the highest reported so far for an integrated quantum device\cite{Zhu2022, Sund2023}, and we operate the device at near gigahertz-level repetition rates, only limited in this demonstration by the repetition rate of our pump laser (\SI{500}{\mega\hertz}). 
Note that the theoretical maximum --- in this device geometry set at \SI{5}{\giga\hertz} --- is only affected by the TB separation, and is in principle only limited by the pulse duration and detector jitter.
With a switching peak-to-peak voltage as low as $V_\pi=\SI{3.37}{\volt}$ at \SI{1}{\mega\hertz} and $V_\pi=\SI{4.35}{\volt}$ at \SI{5}{\giga\hertz}, the device operates at low RF power ($<17$ dBm) and in a CMOS-compatible voltage range, making it suitable for co-integration with standard high-speed electronics.

In terms of visibility of quantum interference, the device reaches a maximum of $\left(87.9 \pm 0.7\right)\%$ without subtraction of accidentals and {$\left(88.9 \pm 0.7\right)\%$} with subtraction of accidentals, when operated in active switching regime.
The former value, which is a more realistic metric for practical field-deployed applications such as QKD, is partially limited by the finite coincidence-to-accidental ratio (CAR), here approximately 60, to a maximum value of $\mathcal{V}_{\rm max}=\mathrm{CAR} / (\mathrm{CAR} + 2) \approx 96.8\%$, and partially by the finite extinction of the interferometers and accuracy in the active switching driving parameters.
While the CAR is mainly a figure of merit pertaining to the SFWM source, the impact of the latter nonidealities can be, in principle, mitigated by reducing the multi-mode and polarization cross-talk in the PIC, which we believe to ultimately limit the visibility. This can be achieved by exploiting Euler bends instead of standard circular curves \cite{Jiang:18} and including polarization filters. Future iterations of the setup could involve further optimization of its components to reduce optical losses, which are especially critical in multi-photon applications.

While here we limit ourselves to showcase the device operation with a demonstration of PSL-free entanglement certification, future works may include its application to quantum communication protocols, such as QKD, teleportation, and entanglement swapping, as well as for the automated tomography of TB qubits.
For all of these applications, the use of TBs would be particularly favorable over fiber (but also free space) optical links, owing to the inherent robustness of the TB encoding over long-haul propagation.
Further investigation may be also devoted to the extension of this approach to higher dimensional TB states (qudits).

In summary, we have demonstrated a fully engineered device for entanglement certification in TFLN integrated photonics.
Operating at gigahertz-rate switching speed, and overcoming the need for temporal post-selection, this result is a significant step toward the practical deployment of quantum communication protocols over actual quantum networks. 

\footnotesize
\vspace{0.2cm}
\noindent\textbf{Funding}\\
M.B, A.B., M.C., and M.G. acknowledge support of the Italian Ministry of Education (MUR) PNRR project PE0000023-NQSTI. 
S.C. acknowledges the HyperSpace project (project ID 101070168).
D.B acknowledges the support of MUR and the European Union - Next Generation EU through the PRIN project number F53D23000550006 - SIGNED.

\vspace{0.2cm}
\noindent\textbf{Acknowledgements}\\
The authors acknowledge CEA-Leti, Grenoble for providing the silicon photonic chip used as source of photon pairs.

\renewcommand{\bibpreamble}{
$^\dagger$These authors contributed equally to this work.\\
$^*${Corresponding author: \\ \textcolor{blue}{marcello.bacchi01@universitadipavia.it}}\\
}

%%%%%%%%%%%%%%%%%%%%%%% References %%%%%%%%%%%%%%%%%%%%%%%%%

\bibliography{references}

\end{document}